\documentstyle[aps,mprocl,psfig]{revtex}
\begin{document}
\title{{\normalsize In: D. E. Wolf, P. Grassberger (Eds.) ``Friction,
Arching, Contact Dynamics'' World Scientific (Singapore, 1997) p. 293-299}\\[1.5cm]
Contact of Viscoelastic Spheres}
\author{Thomas Schwager and Thorsten 
 P\"oschel~\footnote{schwager@summa.physik.hu-berlin.de, 
 thorsten@itp02.physik.hu-berlin.de}}
\address{Humboldt-Universit\"at zu Berlin,
Institut f\"ur Physik, \\
Invalidenstra\ss e 110,
D-10115 Berlin, Germany}
\draft
\date{\today}
\maketitle

\begin{abstract}
  In a recent paper an implicit equation for contacting viscoelastic
  spheres was derived~\cite{BSHP}. Integrating this equation it can be
  shown that the coefficient of normal restitution $\epsilon$ depends
  on the impact velocity $g$ as $1-\epsilon\sim g^{\frac{1}{5}}$
\end{abstract}

The behavior of granular gases has been of large scientific interest
in recent time. Goldhirsch and Zanetti~\cite{Goldhirsch} and McNamara
and Young~\cite{McNamara} have shown that a homogeneous granular gas
is unstable. After some time one observes dense regions (clusters) and
voids. To evaluate the loss of mechanical energy due to collisions one
introduces the coefficient of (normal) restitution
\begin{equation}
  \label{restitution}
  g^\prime = - \epsilon g
\end{equation}
describing the loss of relative normal velocity $g^\prime$ of a pair
of colliding particles after the collision with respect to the impact
velocity $g$

In the investigations~\cite{Goldhirsch,McNamara} the approximation of
constant coefficient of restitution was assumed. Solving viscoelastic
equations for spheres currently it was shown that the coefficient of
normal restitution $\epsilon$ is not a constant but a function of the
impact velocity $\epsilon\left(g\right)$ itself\,\cite{BSHP,KK}. For
the ``compression'' $\xi = R_1+R_2 -\left|\vec{r}_1 - \vec{r}_2\right|
$ of particles with radii $R_1$ and $R_2$ at positions $\vec{r}_1$ and
$\vec{r}_2$ one finds
\begin{equation}
\ddot \xi +\rho\left( \xi^{3/2} +\frac{3}{2}\,A\, 
\sqrt{\xi}\, \dot {\xi}\right)=0
\label{xitaylor}
\end{equation}
\begin{equation}
  \rho=\frac{2~ Y\sqrt{R_{\,\mbox{\it \footnotesize eff}}}} 
  {3~ m_{\mbox{\it \footnotesize eff}}\left( 1-\nu ^2\right) }
\end{equation}
\begin{eqnarray}
  m_{\mbox{\it \footnotesize eff}}=\frac{m_1 m_2}{m_1+m_2} 
  ~~~~~~~ R_{\mbox{\it \footnotesize eff}}=\frac{R_1 R_2}{R_1+R_2} 
  \nonumber
\end{eqnarray}
$Y$ is the Young modulus and $m_{\mbox{\it \footnotesize eff}}$ and
$R_{\mbox{\it \footnotesize eff}}$ are the effective radius and mass
of the grains, respectively. $A$ is a material constant depending on
the Young modulus, the viscous constants and the Poisson ratio of the
material. The initial conditions for solving (\ref{xitaylor}) are
$\xi(0) = 0$ and $\dot{\xi}(0) = g$. The coefficient of restitution
$\epsilon$ of at time $t=0$ colliding spherical grains can be found
from this equation relating the relative normal velocities $g = \dot
\xi(0)$ at time of impact and at time $t_c$, when the particles
separate after the collision, i.e. $t_c$ is the collision time:
\begin{equation}
  \label{eps1}
  \epsilon=-\dot{\xi} \left(t_c\right)/\dot{\xi}\left(0\right).
\end{equation}
The (numerical) integration of equation (\ref{eps1}) yields the
coefficient of restitution as a function of the impact velocity (see
fig.~1 in ref.~\cite{BSHP}) which is in good agreement with
experimental data\cite{Bridges}. Constant coefficient of restitution,
however, does {\em not} agree with experimental
experience~\cite{Hatzes}. Other theoretical work on this topic can be
found e.g. in~\cite{Johnson,pao}.

The duration of collision $t_c^{0}$ for the undamped problem ($A=0$)
is given by~\cite{Hertz:1882}
\begin{equation}
  \label{Thetadef}
  t_c^{0}=\frac{\Theta_c^{0}}{\rho^{\frac{2}{5}}g^{\frac{1}{5}}} ~,
\end{equation}
with $\Theta_c^0$ being a constant. Substituting
$\xi=\rho^{-2}x(\Theta)$ we get for the rescaled velocity
$v=\rho^{2}g$, and with (\ref{Thetadef}) one finds
$t=\Theta\,v^{-\frac{1}{5}}$. We rewrite (\ref{xitaylor}) using
rescaled compression, relative velocity and time $x,v$ and $\Theta$
and the abbreviation $\alpha=\frac{3}{2}A$
\begin{equation}
  \label{initial}
  \ddot{x} + \alpha v^{-\frac{1}{5}} \dot{x}\sqrt{x} + 
  v^{-\frac{2}{5}}x^{\frac{3}{2}} = 0
\end{equation}
with $\dot{x}=\frac{d}{d\Theta}x$. The initial conditions read now 
\begin{eqnarray}
  x(0) &=& 0\\
 \frac{dx}{dt}(0)&=& v^{\frac{1}{5}}\frac{dx}{d\Theta}(0)=v, 
 ~~\mbox{hence}~~\dot{x}(0) = v^{\frac{4}{5}}.
\end{eqnarray}
Eq.~(\ref{initial}) will be solved now by series expansion. All
derivatives of third order and higher of $x$ diverge at $\Theta=0$,
hence one cannot expand $x$ in powers of $\Theta$.  Since $\xi$ can be
written in the form $\xi(t)=g\,t\,y(t)$, the Ansatz
 \begin{equation}
   x(\Theta)=v^{\frac{4}{5}}\Theta~(1+\eta(\Theta))~~~
   \mbox{with~~}\eta(0)=0
 \end{equation}
seems to be a reasonable assumption. We get an equation for $\eta$ :
\begin{equation}
  \Theta\ddot{\eta} + 2 \dot{\eta} + \alpha v^{\frac{1}{5}} 
  \Theta^{\frac{3}{2}}\dot{\eta}\sqrt{1+\eta} + 
  \left(\alpha v^{\frac{1}{5}}\sqrt{\Theta} + 
    \Theta^{\frac{3}{2}}\right) (1+\eta)^{\frac{3}{2}} = 0~.
\label{etaeq}
\end{equation}
In (\ref{etaeq}) occur terms $\Theta^{0.5}$ and $\Theta^{1.5}$,
therefore we expand $\eta$ in powers of $\sqrt{\Theta}$
\begin{equation}
  \label{etaseries}
  \eta=\sum_{k=0}^\infty a_k~\Theta^{\frac{k}{2}}
\end{equation}
The first coefficients $a_0=0$ and $a_1=0$ vanish because of the
initial conditions. With Taylor expansion of $\sqrt{1+\eta}$ and
$(1+\eta)^{\frac{3}{2}}$ for small $\eta$ we arrive at
\begin{equation}
  \label{firstTermseta}
  \eta = -\frac{4}{15} \alpha v^{\frac{1}{5}}  \Theta^\frac{3}{2} 
  - \frac{4}{35} \Theta^{\frac{5}{2}} + \frac{3}{70} 
  \alpha v^{\frac{1}{5}} \Theta^4 + \frac{1}{15} \alpha^2 
  v^{\frac{2}{5}} \Theta^3 \dots
\end{equation}
and therefore
\begin{eqnarray}
\label{firstTermsx}
  x &=& v^{\frac{4}{5}}~\Theta - \frac{4}{15}\alpha 
  v~\Theta^{\frac{5}{2}} - \frac{4}{35} v^{\frac{4}{5}} 
  \Theta^{\frac{7}{2}} + \frac{1}{15} \alpha^2 v^{\frac{6}{5}} 
  \Theta^4\nonumber \\
  && + \frac{3}{70} \alpha v \Theta^5 - \frac{38}{2475} \alpha^3 
  v^{\frac{7}{5}} \Theta^{\frac{11}{2}} + 
  \frac{1}{175} v^{\frac{4}{5}} \Theta^6 + \dots
\end{eqnarray}

\begin{figure}[htbp]
\vspace{-0.2cm}
\centerline{\psfig{figure=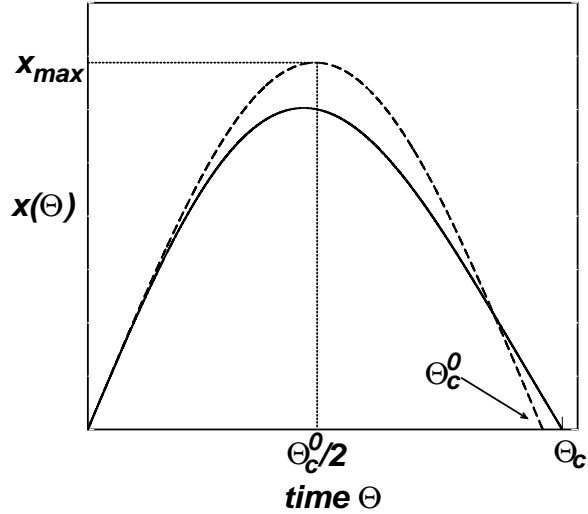,width=9cm,angle=0}}
\vspace{-0.5cm}
\caption{The dynamics of the collision. The dashed line shows 
  the (strictly symmetric) solution of the undamped collision. For the
  case of the damped motion (full line) the maximum penetration depth
  is achieved earlier whereas the duration of the collision is longer
  ($\Theta_c>\Theta_c^0$). The figure gives a nice impress why we
  employ the {\em inverse collision} instead of direct calculation:
  one cannot expand $x$ beyond $\Theta_c^0$. In the point $\Theta_c^0$
  the derivatives of the curve for elastic motion diverge.}
\label{fig:trajectory}
\end{figure}

\noindent Rearranging the full series (\ref{firstTermsx}) one 
finds (s.~fig.~\ref{fig:trajectory})
\begin{equation}
  x = v^{\frac{4}{5}}x_0(\Theta) + \alpha v x_1(\Theta) + \alpha^2 
  v^{\frac{6}{5}} x_2(\Theta) \dots \label{xkdef} 
\end{equation}
$v^{\frac{4}{5}}x_0$ is the solution of the undamped (elastic)
collision. We will need $x\left(\frac{1}{2}\Theta_c^0\right)$ where
$\Theta_c^0$ is the duration of the undamped collision.

Later we will need the solution $x^{inv}$ of the {\em inverse
  problem}, i.e. of a collision with impact velocity $v^\prime$ and
final velocity $v$. Hence the inverse collision is not a damped
motion, but an accelerated one, where $\alpha$ has to be replaced by
$-\alpha$. Substituting $v\rightarrow v^\prime$ and $\alpha\rightarrow
-\alpha$ we find
\begin{equation}
  \label{xinv}
  x^{inv}(\Theta') = (v^\prime)^{\frac{4}{5}}x_0(\Theta') - \alpha 
  v^\prime x_1(\Theta') + \alpha^2 (v^\prime)^{\frac{6}{5}} 
  x_2(\Theta') \dots
\end{equation}
Now we determine the collision time $\Theta_c$ and the final velocity
$\frac{d}{dt}x\left(\Theta_c\right)$. (A more direct method to
calculate $\Theta_c$ is to determine the solution of $x(\Theta) = 0$
using Taylor expansion of $x$ in the region close to $\Theta_c^0$. It
can be seen easily that this method fails since, all derivatives of
$\frac{\partial^n x}{\partial\Theta^n}$ with $n\ge 3$ diverge for
$\Theta=\Theta_c^0$. Therefore $\Theta_c$ has to be calculated
indirectly as explained in the text (s.~fig.~\ref{fig:trajectory}).)

\begin{figure}[hbp]
  \centerline{\psfig{figure=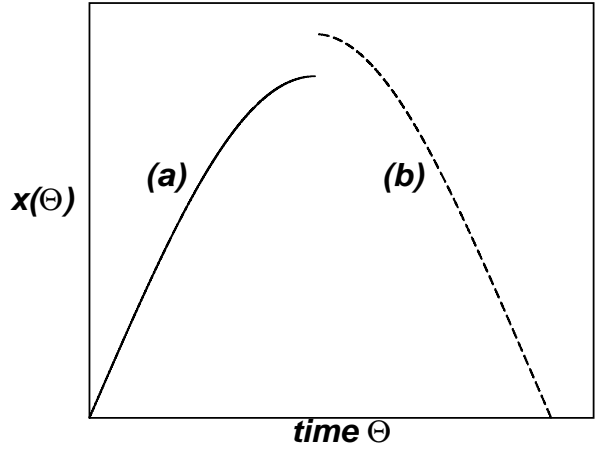,width=9cm,angle=0}}
  \vspace{-0.7cm}
  \caption{Sketch of the calculation. The first part $(a)$ 
    $\Theta\in\left(0,\Theta_m\right)$ is calculated directly, for the
    other part $(b)$ we define the {\em inverse collision} where the
    particles start with velocity $v^\prime$ and velocity approaches
    zero at $\Theta=\Theta_m$. Both curves have to fit together
    smoothly.}
\label{fig:sketch}
\end{figure}

The problem will be subdivided into two parts (s.
fig.~\ref{fig:sketch}): a) the motion of the particles $x$ from
$\Theta=0$ to time $\Theta_m$ when $x$ approaches its maximum and
where $\dot{x}$ changes its sign, and b) from $\Theta_m$ to
$\Theta_c$. In case of undamped motion where $\alpha=0$ we have
$\Theta_m = \Theta_c^0/2$. In part b) we do not consider the collision
itself but the inverse problem in the interval
$(\Theta=0,~\Theta_m^\prime)$, with $\Theta_m^\prime$ being the time
where $x^{inv}$ approaches its maximum.  The continuity of both parts
means
\begin{equation}
  x\left(\Theta_m\right) = x^{inv}\left(\Theta_m^\prime\right)~.
\end{equation}

\begin{figure}[htbp]
\vspace{-0.5cm}
    \centerline{\psfig{figure=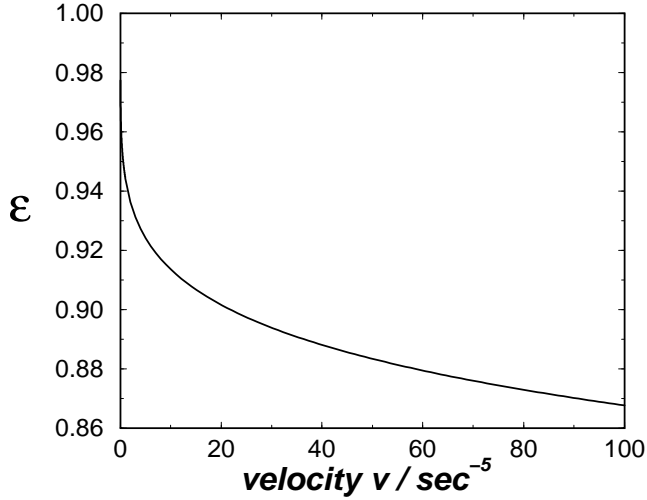,width=9cm,angle=0}}
\vspace{-0.5cm}
\caption{The coefficient of restitution over impact velocity due 
  to eq. (\ref{epsfinal}). As expected for small relative velocity the
  particles collide almost elastically. The result of numerical
  integration of (\ref{eps1}) coincides with the curve. Both curves
  cannot be distinguished in the plot.}
\label{fig:ev}
\end{figure}

For finite damping $\alpha\ne 0$ we write $\Theta_m = \Theta_c^0/2 +
\delta$ and $\Theta^\prime_m = \left(\Theta_c^0\right)^\prime/2 +
\delta^\prime$ and remind that $\Theta_c^0 =
\left(\Theta_c^0\right)^\prime $ .  To get an expression for $\delta$
we expand
\begin{eqnarray}
  \dot{x}\left(\frac{\Theta_c^0}{2} + \delta \right) = 0 &=& 
  \dot{x}\left(\frac{\Theta_c^0}{2}\right) + \delta 
  \ddot{x}\left(\frac{\Theta_c^0}{2}\right) + \frac{\delta^2}{2} 
  \frac{d^3}{d\Theta^3} x\left(\frac{\Theta_c^0}{2}\right) + \dots \\
  &=& v^{\frac{4}{5}}\left(\dot{x}_0\left(\frac{\Theta_c^0}{2}\right) + 
    \delta \ddot{x}_0\left(\frac{\Theta_c^0}{2}\right) + 
    \frac{\delta^2}{2} \frac{d^3}{d\Theta^3} x_0\left( 
      \frac{\Theta_c^0}{2}\right) + \dots\right) \\
  &&+ v\alpha\left( \dot{x}_1\left(\frac{\Theta_c^0}{2}\right) + 
    \delta \ddot{x}_1\left(\frac{\Theta_c^0}{2}\right) + \frac{\delta^2}{2} 
    \frac{d^3}{d\Theta^3} x_1\left(\frac{\Theta_c^0}{2}\right)  + 
    \dots\right)\nonumber \\
\end{eqnarray}
and using $\dot{x}_0\left(\Theta_c^0/2\right) = 0$
($v^\frac{4}{5}~x_0$ is the solution of the undamped problem)
\begin{equation}
  \delta = -\alpha v^{\frac{1}{5}} \frac{\dot{x}_1\left(\Theta_c^0/2\right)}
  {\ddot{x}_0\left(\Theta_c^0/2\right)} + {\cal O} \left(\alpha^{2}\right)~.
\label{delta}
\end{equation}
The expression (\ref{delta}) has to be inserted into the Taylor expansion 
of $x\left(\Theta_c^0/2 + \delta \right)$:
\begin{eqnarray}
x\left(\Theta_c^0/2 + \delta \right) 
& =& v^{\frac{4}{5}}\left(x_0\left(\frac{\Theta_c^0}{2}\right) + 
  \delta \dot{x}_0\left(\frac{\Theta_c^0}{2}\right) + \frac{\delta^2}{2} 
  \ddot{x}_0\left(\frac{\Theta_c^0}{2}\right) + \dots\right)\nonumber \\
&&~~ + \alpha v \left(x_1\left(\frac{\Theta_c^0}{2}\right) + \delta 
  \dot{x}_1\left(\frac{\Theta_c^0}{2}\right) + \frac{\delta^2}{2} 
  \ddot{x}_1\left(\frac{\Theta_c^0}{2}\right)  + \dots\right)\\
&& = v^{\frac{4}{5}}x_0\left(\frac{\Theta_c^0}{2}\right) + \alpha 
v x_1\left(\frac{\Theta_c^0}{2}\right) - \frac{\alpha^2 v^{\frac{6}{5}}}{2} 
\frac{\dot{x}_1^2\left(\Theta_c^0/2\right)}{\ddot{x}_0\left(\Theta_c^0/2\right)} 
\nonumber\\
&&~~ + \alpha^2 v^{\frac{6}{5}} x_2\left(\frac{\Theta_c^0}{2}\right) + 
{\cal O}\left(\alpha^3\right)~.
\label{xmax1}
\end{eqnarray}
Hence
\begin{eqnarray}
  x\left(\Theta_m\right) &=& v^{\frac{4}{5}} x_0\left(\Theta_c^0/2\right) + 
  \alpha v x_1\left(\Theta_c^0/2\right) + \nonumber\\
&&~~~~~~\alpha^2 v^{\frac{6}{5}} \left(x_2\left(\frac{\Theta_c^0}{2}\right)- 
  \frac{1}{2}\frac{\dot{x}_1^2\left(\Theta_c^0/2\right)}
  {\ddot{x}_0\left(\Theta_c^0/2\right)}\right)+\dots 
\label{sol1}
\end{eqnarray}
Replacing again $v\rightarrow v^\prime$ and $\alpha\rightarrow -\alpha$ yields
\begin{eqnarray}
  \delta^\prime &=& \alpha (v^\prime)^{\frac{1}{5}} \frac{\dot{x}_1\left(
      \Theta_c^0/2\right)}{\ddot{x}_0\left(\Theta_c^0/2\right)} + {\cal O} 
  \left(\alpha^{2}\right)\\
  \label{deltas}
 x^{inv}\left(\Theta_m^\prime\right) &=& (v^\prime)^{\frac{4}{5}} 
 x_0\left(\Theta_c^0/2\right) - \alpha v^\prime x_1\left(\Theta_c^0/2\right) + 
 \nonumber\\
 &&~~~~~~~~~~~~~~~\alpha^2 (v^\prime)^{\frac{6}{5}} 
 \left(x_2\left(\frac{\Theta_c^0}{2}\right)- \frac{1}{2}
   \frac{\dot{x}_1^2\left(\Theta_c^0/2\right)}{\ddot{x}_0
     \left(\Theta_c^0/2\right)}\right)+\dots
  \label{sol2}
\end{eqnarray}
As explained above both solutions (\ref{sol1}) and (\ref{sol2}) have 
to be equal. With 
\begin{equation} 
  \beta = x_2\left(\frac{\Theta_c^0}{2}\right)- \frac{1}{2}
  \frac{\dot{x}_1^2\left(\Theta_c^0/2\right)}{\ddot{x}_0
    \left(\Theta_c^0/2\right)}
\end{equation}
we write
\begin{eqnarray}
v^{\frac{4}{5}} x_0\left(\frac{\Theta_0^c}{2}\right) + \alpha 
v x_1\left(\frac{\Theta_0^c}{2}\right) + \alpha^2 v^{\frac{6}{5}}\beta 
&=& (v^\prime)^{\frac{4}{5}} x_0\left(\frac{\Theta_0^c}{2}\right) - 
\alpha v^\prime  x_1\left(\frac{\Theta_0^c}{2}\right)\nonumber\\
&&~~~ + \alpha^2 (v^\prime)^{\frac{6}{5}}\beta ~. 
\end{eqnarray}
We expand $v^\prime$ in $\alpha$ 
\begin{equation}
  v^\prime = v + \alpha v_1 + \alpha ^2 v_2 + \dots ~,
\end{equation}
and find
\begin{eqnarray}
&&  v^{\frac{4}{5}} x_0\left(\frac{\Theta_0^c}{2}\right) + \alpha 
v x_1\left(\frac{\Theta_0^c}{2}\right) + \alpha^2 v^{\frac{6}{5}}\beta 
\nonumber\\
&=& v^{\frac{4}{5}}\left(1+ \frac{\delta v}{v}\right)^{\frac{4}{5}} 
x_0\left(\frac{\Theta_0^c}{2}\right) - \alpha v\left(1 + 
  \frac{\delta v}{v}\right) x_1\left(\frac{\Theta_0^c}{2}\right) + 
\alpha^2 v^{\frac{6}{5}}\left(1 + \frac{\delta v}{v}\right)^{\frac{6}{5}} 
\beta 
\end{eqnarray}
with $\delta v = \alpha v_1 + \alpha^2 v_2 + \dots$. Writing 
$(1+\frac{\delta v}{v})^{\frac{n}{5}}$ in powers of $\alpha$ and 
comparing coefficients yields finally
\begin{eqnarray}
  \label{solvs}
  v^\prime &=& v \left(1+\frac{5}{2}\alpha v^{\frac{1}{5}}
    \frac{x_1\left(\frac{\Theta_c^0}{2}\right)}{x_0\left(
        \frac{\Theta_c^0}{2}\right)} + 
    \frac{15}{4}\alpha^2 v^{\frac{2}{5}}\left(
      \frac{x_1\left(\frac{\Theta_c^0}{2}\right)}{x_0\left(
          \frac{\Theta_c^0}{2}\right)}\right)^2 
    + \dots \right)\nonumber \\
  &=&  v \left(1-\alpha v^{\frac{1}{5}} C_1 + \alpha^2 
    v^{\frac{2}{5}} C_2 + \dots\right)
\label{Cdef}
\end{eqnarray}
and for coefficient of normal restitution one gets
\begin{eqnarray}
  \epsilon &=& \frac{v^\prime}{v} = 1-\alpha v^{\frac{1}{5}} C_1 + 
  \alpha^2 v^{\frac{2}{5}} C_2 + \dots\\
  &=& 1-\frac{3}{2}C_1 A\rho^{\frac{2}{5}}g^{\frac{1}{5}} + 
  \frac{9}{4}C_2 A^2\rho^{\frac{4}{5}}g^{\frac{2}{5}} + \dots 
\label{epsfinal}
\end{eqnarray}
with $g$ being the impact velocity~(fig~\ref{fig:ev}). For the
duration of collision we find with (\ref{delta}), (\ref{deltas}) and
(\ref{solvs})
\begin{equation}
  t_c = \Theta_c^0 \rho^{-\frac{2}{5}}g~^{-\frac{1}{5}}
  \left(1+\frac{1}{10}C_1\alpha \rho^{\frac{2}{5}}
    g^{\frac{1}{5}}\right) + {\cal O}(\alpha^2)~.
\end{equation}

Our final results eq.~(\ref{epsfinal}) shows that for viscoelastic 
colliding smooth bodies the coefficient of normal restitution is a 
decreasing function with rising impact velocity: 
$1-\epsilon \sim g^\frac{1}{5}$.

The authors are grateful to N.~Brilliantov, S.~Esipov, H.~Herrmann,
F.~Spahn and W.~Young for helpful discussions, and J.-M. Hertzsch for
providing relevant literature.

\end{document}